\documentstyle[preprint,aps,epsfig]{revtex}

\begin{document}

\draft
\tighten
\preprint{
\vbox{
\hbox{ADP-99-31/T368}
\hbox{IU/NTC 99-01}
}}

\title{Structure and Production of Lambda Baryons }

\normalsize
\author{C. Boros, J.T. Londergan$^1$ and  A.W. Thomas}
\address {Department of Physics and Mathematical Physics,
                and Special Research Center for the
                Subatomic Structure of Matter,
                University of Adelaide,
                Adelaide 5005, Australia}

\address{$^1$ Department of Physics and Nuclear
            Theory Center, Indiana University,
            Bloomington, IN 47408, USA}

\date{\today}
\maketitle

\begin{abstract} 
We discuss the quark parton structure of the $\Lambda$ 
baryon and the fragmentation of quarks into $\Lambda$ baryons. 
We show that 
the hyperfine interaction, responsible for the 
$\Delta$-$N$ and  $\Sigma^0$-$\Lambda$ mass splittings,  
leads not only to sizeable $SU(3)$ and 
$SU(6)$ symmetry breaking in the quark distributions of the $\Lambda$,   
but also to significant polarized non-strange quark distributions.   
The same arguments  
suggest flavor asymmetric quark fragmentation functions  
and non-zero polarized non-strange quark fragmentation functions.  
The calculated fragmentation functions give a good description of 
all measured observables.  
We predict significant positive $\Lambda$ polarization  
in semi-inclusive DIS experiments  
while models based on SU(3) flavor symmetry 
predict zero or negative $\Lambda$ polarization.  
Our approach also provides  a natural explanation for the dependence 
of the maximum of the 
$\xi=\ln(1/z)$ spectrum  on the mass of the particles  produced 
in $e^+e^-$ annihilation.  

\end{abstract} 

\newpage

\section{Introduction} 

An impressive amount of information on the quark parton structure 
of the nucleons has been collected since the pioneering 
experiment at SLAC which showed the first evidence of nucleon partonic 
substructure \cite{Friedman}.  
However, significantly less is known 
about the structure of other baryons. This is because of
the impossibility of producing targets of short lived baryons  
for lepton nucleon deep-inelastic scattering [DIS] experiments 
which might measure their  
unpolarized or polarized structure functions. 
One possibility is to measure the fragmentation 
functions of quarks into the baryons and relate the information 
obtained in these experiments to the quark structure of the 
baryons. The Lambda hyperon is of special interest in this 
respect since its decay is self-analyzing.  Polarization measurements 
are thus relatively simple to perform and the polarized 
fragmentation functions of quarks into $\Lambda$ can be 
measured.  Furthermore, in the quark parton model the 
$\Lambda$ has a rather simple structure: the $u$ and $d$ quarks  
couple to a spin and an isospin singlet state, so the $\Lambda$ 
spin is carried exclusively by its strange quark.     

As is well known,  
the naive quark model fails to explain the 
data on hyperon $\beta$ decay and 
on deep inelastic scattering \cite{EMC,E143,SMC,Hermes}. 
The violation of the Ellis-Jaffe sum rule \cite{Ellis} 
suggests a large strange quark polarization in the 
nucleon.  These observations suggest 
that the non-strange 
quarks of the $\Lambda$ might also be substantially polarized 
\cite{Jaffe93}. This and related questions, together 
with the experimental feasibility of $\Lambda$ 
polarization experiments, have stimulated much theoretical activity 
\cite{Jaffe93}-\cite{Ma} on this subject.   
Information on the   
structure of the $\Lambda$ should lead eventually to a deeper 
understanding of the structure of the nucleon.

In this paper, we re-examine assumptions such as $SU(3)$ flavor 
and $SU(6)$ spin-flavor symmetry, which are frequently crucial 
elements in predictions of $\Lambda$ baryon structure. 
In Sect.\ II we point out that one should expect both the unpolarized 
and polarized
quark distributions in the Lambda to show substantial differences
from predictions based on either  
$SU(3)$ flavor or $SU(6)$ spin-flavor symmetry.  
The same mechanism which is responsible 
for breaking the SU(6) symmetry of 
the nucleon's quark distributions leads to 
$x$-dependent, polarized, non-strange valence 
quark distributions in the $\Lambda$.  This contrasts with 
the naive expectation that the up and down valence quarks 
of the $\Lambda$ should be unpolarized.  
In order to estimate the magnitude of these symmetry 
breaking effects and the size of the up and down quark polarizations,   
the quark distributions in the $\Lambda$ are calculated in the 
MIT bag model.  
In Sect.\ III, we discuss how 
these flavor symmetry breaking effects carry over into the  
fragmentation of quarks into $\Lambda$ baryons. 
In Sect.\ IV.A, we show that the calculated 
fragmentation functions give an good overall description of all 
measured observables in inclusive particle production in $e^+e^-$ 
annihilation.  We also discuss the relevance of our 
approach to the dependence of the maximum of the 
$\xi=\ln(1/z)$ spectrum  on the mass of the produced particles.   
In Sect.\ IV.B it is demonstrated that these polarized non-strange 
quark distributions give rise to sizeable $\Lambda$ polarization 
in polarized semi-inclusive DIS experiments,  
in contrast to predictions based on $SU(3)$ flavor symmetry. 

\section{Quark distributions in the $\Lambda$} 

Baryon flavor symmetry is widely used to relate 
the structure of particles within the baryon octet.    
Flavor symmetry breaking effects are generally  
accounted for by using different phenomenological  
masses for the strange quark than for up and down 
quarks. However, $SU(3)$ symmetry breaking can be much more 
subtle, as has been pointed out in Ref. \cite{Alberg,Boros99}.   
In a world of exact $SU(6)$ spin-flavor symmetry 
the up and down quark distributions should be identical.  
Experiments indicate that they are different from each other; for 
example, it is known that the ratio $d(x)/u(x)$ drops 
rapidly below unity as $x\rightarrow 1$. 
The hyperfine interaction responsible for the splitting 
for the $\Delta$-$N$ masses gives a natural explanation 
for this observation \cite{Close73,Carlitz,Close88}.   
It is not yet clear whether this ratio actually goes to zero at $x=1$ 
or approaches the pQCD limit of $0.2$ \cite{pQCD1,pQCD2}, 
although the latter now seems favoured \cite{Melni96,Yang99}.  
If $SU(3)$ flavor symmetry is used to relate the $\Sigma^+$ 
quark distributions to those in the proton, one would  
predict an analogous large-$x$ behavior,  
$s_\Sigma/u_\Sigma\rightarrow 0$ for $x\rightarrow 1$.  However, 
as has been pointed out in Ref. \cite{Alberg,Boros99}, the 
hyperfine interaction responsible for the splitting
of the $\Sigma$-$\Lambda$ masses predicts  
a behavior opposite to the $SU(3)$ expectation. 
For this reason it is mandatory that we re-examine  
$SU(3)$ symmetry arguments in the $\Lambda$ case. 
This has been done partly in our previous 
paper on the quark distributions 
in the $\Lambda$ \cite{Boros99} and   
in a quark diquark spectator model for the fragmentation functions 
in Ref.\cite{Hood95} and also for the quark distribution fucntions 
\cite{Jacob}. Here, we extend our earlier discussion 
to fragmentation, emphasizing  the close relationship  
between distribution and fragmentation functions.  
Since the publication of our paper \cite{Boros99}, 
there have been discussions along similar lines by Ma 
{\it et al.} \cite{Ma}.

It is instructive to review how the QCD hyperfine interaction 
breaks $SU(6)$ spin-flavor symmetry.   
The leading -twist quark distributions
can be formally defined as \cite{Collins,Jaffe83}
\begin{equation}
  q_\Gamma(x) = P^+ \int \frac{d\xi^-}{2\pi}
      e^{ixP^+ \xi^-}
   \langle \Lambda ; PS   | \overline{\psi} (0)
  \Gamma \psi (\xi^-) | \Lambda ; PS\rangle , 
\end{equation}
where $\Gamma$ is a Dirac matrix; $P$ and $S$ are respectively the 
momentum and spin of the 
$\Lambda$ and we defined $P^\pm\equiv P^0\pm P^3$. 
Inserting a complete set of intermediate states,
using the translation invariance of the matrix elements
and the integral representation of the $\delta$-function 
the twist-two, 
helicity projections, $q^{\uparrow \downarrow}$,   
are given by  
\begin{equation}
  q^{\uparrow \downarrow} (x) 
 = 2 P^+  \sum_n \delta [(1-x)P^+ -p_n^+]
 \,\, | \langle n ; p_n  | 
\psi_+^{\uparrow\downarrow} (0) | \Lambda ; PS_\parallel \rangle |^2  . 
\label{eq2} 
\end{equation} 
Here, $\psi_+^{\uparrow\downarrow} =\frac{1}{2}(1+\gamma_0\gamma_3)
\frac{1}{2}(1\pm \gamma_5)\psi$ with    
$P_+=\frac{1}{2}(1+\gamma_0\gamma_3)$ and 
$\Lambda_\pm=\frac{1}{2}(1\pm \gamma_5)$  
(=$\frac{1}{2}(1\mp \gamma_5)$)  
are the relevant light-cone   helicity projection operators 
for quarks (antiquarks) and we defined  $\gamma^+ = \gamma^0+\gamma^+$. 
$S_\parallel$ is the spin vector parallel to the target's three-momentum. 
Further, the states $|n; p_n\rangle $  
are intermediate states with mass $M_n$ and form a
complete set of states with $p_n^+ = \sqrt{M_n^2+{\bf p}_n^2}
+p_{nz}$. All states  are normalized 
to $(2\pi)^3 \delta ({\bf p} - {\bf p^\prime})$.  
$q^{\uparrow\downarrow}$ can be interpreted as the probability 
to find a quark with the same/opposite helicity as the target 
hyperon. 

The advantage of using Eq.(\ref{eq2}) is that energy-momentum 
conservation is ensured so that the resulting quark 
distributions have correct
support.  This is guaranteed by Eq.(\ref{eq2}) regardless
of the approximation used for the states $|n; {\bf p_n}\rangle$ and
$|\Lambda;PS\rangle$.       
The quark operator $\psi$ acts on the initial state $|\Lambda;PS\rangle$. 
It  either destroys a quark, producing an intermediate two quark state,  
or it inserts an anti-quark in the target, producing a four 
quark intermediate state.  The delta function implies
that the contribution to the quark distribution arising from 
an intermediate state with mass $M_n$ peaks at 
\begin{equation}
x_{max} \approx (1-M_n/M_\Lambda) ~~.
\label{xpeak}
\end{equation}  
As a result, the shape of the quark 
distribution at large $x$ is determined by the smallest mass $M_n$
which can contribute to the particular distribution.  
Since the mass of the intermediate state, in general, depends 
on the flavor of the struck quark, this mass dependence translates 
into a flavor dependence of the quark distribution functions 
of the baryon. We also see that contributions from four-quark intermediate 
states peak at {\it negative} x values, since $M_n>M$, and are thus    
suppressed in the positive $x$ region.

QCD color-magnetic effects lift the mass degeneracy between hadrons 
that differ only in the 
orientation of quark spins, such as $N$ and $\Delta$.  
The interaction is repulsive if the spins are parallel, so that a 
pair of quarks in a spin-1 state (vector) has higher energy than a pair 
of quarks in a spin-0 state (scalar).   
The energy shift between scalar and vector diquarks produces 
the $N-\Delta$ mass splitting.  
The $\vec{s}_i \vec{s}_j$ structure of the hyperfine 
interaction shifts the mass of the vector and scalar diquarks 
in the ratio $1:-3$. From the experimental $\Delta$-$N$ mass 
difference, we conclude that the triplet diquark is heavier by $50$ 
MeV than the diquark state without hyperfine interaction, while the 
singlet diquark is lighter by $150$ MeV.  The diquark masses which 
reproduce the $N$ and $\Delta$ masses are roughly $m_s \approx 600$ 
MeV and $m_v \approx 800$ MeV for the scalar and vector diquarks, 
respectively.  Since the $d$-quark in the proton is always accompanied 
by a vector diquark as opposed to the $u$ quark which has a large 
probability to be accompanied by a scalar diquark,   
the $u$-quark distribution peaks at larger $x$-values than the
$d$-quark distribution.  Using Eq.\ (\ref{xpeak}) with the 
scalar and vector diquark masses, one obtains   
quantitative predictions for the location of the peak in 
the $u$ and $d$ valence quark distributions.

The same arguments applied to the $\Lambda$ and $\Sigma$ mass splitting 
predict that the $us$ vector diquark is heavier by $\approx 30$ MeV, 
and the corresponding scalar diquark is
lighter by $\approx 90$ MeV, than the diquark without
hyperfine splitting. To estimate the masses of diquarks containing a  
strange quark and an up or down quark we use the phenomenological 
fact that the strange quark  
adds about $180$ MeV. Thus, we have $m_s^\prime=800+180-90\approx 890$ MeV 
and $m_v^\prime=800+180+30 =1010$ MeV for singlet and triplet diquarks.

If the struck quark is accompanied by a 
scalar (vector) diquark its distribution peaks at higher (lower) 
$x$ values.  The probabilities for finding a $u$, $d$ or $s$ quark 
polarized parallel or anti-parallel to a $\Lambda$ hyperon, and 
accompanied by a scalar or vector diquark, can be obtained 
from the $SU(6)$ wave function of the $\Lambda$ 
and are given in Table \ref{tab:1}. 
If the struck quark is a strange quark the  
intermediate state must always be a scalar diquark; Eq.\ (\ref{xpeak}) 
shows that this will produce a very hard strange quark distribution in 
the $\Lambda$.  However, if the struck quark is an up or down quark 
the remaining diquark has a higher probability to be a vector diquark 
than a scalar diquark.  This leads  
to softer up and down quark distributions.    

Furthermore, while the valence  $u_v$ or $d_v$ quarks  
with spin anti-parallel to the $\Lambda$ spin are always  
associated with a vector diquark, 
$u_v$ and $d_v$ quarks with spin parallel to the $\Lambda$ spin  
have equal probabilities to be  
accompanied by a vector or scalar diquark. This has the 
important consequence that the distribution of non-strange quarks  
with spin parallel to the $\Lambda$ spin are harder than   
the corresponding distributions with anti-parallel spins. Thus,  
$u_v^\uparrow(x)$ ($d_v^\uparrow(x)$) and 
$u_v^\downarrow(x)$ ($d_v^\downarrow(x)$) are shifted in $x$ relative 
to each other, so that 
$\Delta u_v(x)\equiv u_v^\uparrow (x) -u_v^\downarrow (x)$ and   
$\Delta d_v(x)\equiv d_v^\uparrow (x) -d_v^\downarrow (x)$  
are non-vanishing functions of $x$. They are positive 
for large $x$ and negative for small $x$ values. 
Note that their total contribution to the spin of the $\Lambda$ 
is zero since the integrals over $\Delta u_v$ and 
$\Delta d_v$ are zero. Nevertheless, $\Delta u_v$ and $\Delta d_v$ 
can be sizable for large $x$ values since both $u_v$ and $\Delta u_v$ are 
dominated by the spin-zero component in the large $x$ limit.

These properties of the quark distributions are quite 
general. Once we assume that the intermediate states can be regarded 
as on shell physical states with definite masses,  they follow 
immediately from the definition of the quark distributions and from the 
$SU(6)$ structure of the baryon wave functions.   
Since, up to now, 
quark distributions cannot be calculated from first principles,  
we have to use model wave functions to estimate the magnitude of 
the expected symmetry breaking effects. 
We use MIT bag wave functions and the Peierls-Yoccoz method
for constructing translationally invariant momentum eigenstates,  
$|B_{\{n\}}{\bf p}\rangle$,  from   
$n$ particle bag  states $|B_{\{n\}}({\bf r})\rangle$,
centered at ${\bf r}$ 
\begin{equation}
   |B_{\{n\}}{\bf p}\rangle ,  
 = [\phi_n({\bf p})]^{-1}
   \int d{\bf r} e^{i{\bf pr}}  | B_{\{n\}}({\bf r})\rangle .
\end{equation}
The normalization 
$\phi_n({\bf p})$ is given by
\begin{equation}
|\phi_n({\bf p})|^2 = \int d{\bf R}   e^{-i{\bf p R}} \langle B_{\{n\}}({\bf R})
 |B_{\{n\}}({\bf 0}) \rangle .
\end{equation} 
The matrix element in the  definitions 
of the quark distribution functions can be obtained by 
using these bag states and 
the bag operator $\psi = \sum_m \{ b_m \psi_m(+)  
+ d^\dagger_m \psi (-)\}$ ($\psi_m (\pm)$ are the positive and  
negative energy  
solutions of the Dirac equation in the bag and 
$b_m$, $d^\dagger_m$ are annihilation and creation operators).   
The  spin-dependent parton distributions are then given by 
\cite{Signal,Schreiber}   
\begin{equation} 
q_f^{\uparrow\downarrow}(x) =
  \frac{2 M_\Lambda}{(2\pi )^2} \sum_{m} \langle 
  \Lambda^\uparrow |P_{f,m}|\Lambda^\uparrow   \rangle
\int^\infty_{|\frac{M_\Lambda^2(1-x)^2-M_n^2}{2M_\Lambda(1-x)}|}
p_ndp_n \frac{|\phi_2({\bf p_n})|^2}{|\phi_3({\bf 0})|^2}
 |\psi^{\uparrow\downarrow}_m({\bf p_n})|^2.
\label{eq:dist}
\end{equation} 
Here, $\uparrow\downarrow$ indicates the helicity projections.  
The operator $P_{f,m}$ projects out the appropriate 
spin and isospin  quantum numbers 
from the $SU(6)$ wave functions of the polarized  target baryon. 
Its matrix elements are given in Table \ref{tab:1}.  
$\psi^{\uparrow\downarrow}_m({\bf p_n})$ is the 
Fourier transform of the bag wave function with  
angular momentum component $m$, and may be split 
into spin dependent and spin independent parts, 
\begin{equation}
   |\Psi_m^{\uparrow\downarrow}({\bf p_n})|^2 =
  \frac{1}{2} \left[f({\bf p_n})\pm(-1)^{m+3/2}
  g({\bf p_n}) \right].
\end{equation}

Denoting by $F(x)$ and $G(x)$ those contributions to Eq.
(\ref{eq:dist}) which come from the $f( {\bf p_n})$ and
$g({\bf p_n})$ parts of the integral 
we obtain for the unpolarized distributions 
\begin{eqnarray} 
   d_\Lambda (x) & = & 
 u_\Lambda (x) = \frac{1}{4}[3F_v(x)+F_s(x)]
\nonumber \\ 
 s_\Lambda (x) & =&  F_s(x) ,  
\end{eqnarray} 
and for the polarized quark distribution functions, 
$\Delta q\equiv q^\uparrow - q^\downarrow$, 
\begin{eqnarray} 
\Delta d_\Lambda (x) & = & \Delta u_\Lambda (x)  = 
 \frac{1}{4}[G_s(x) - G_v(x)] \nonumber \\ 
\Delta  s_\Lambda (x) & = & G_s(x),  
\end{eqnarray} 
respectively  
(see Ref .\cite{Signal,Schreiber} 
for details for the proton case and Ref.\cite{Boros99} for other 
baryons).    

\begin{figure}
\psfig{figure=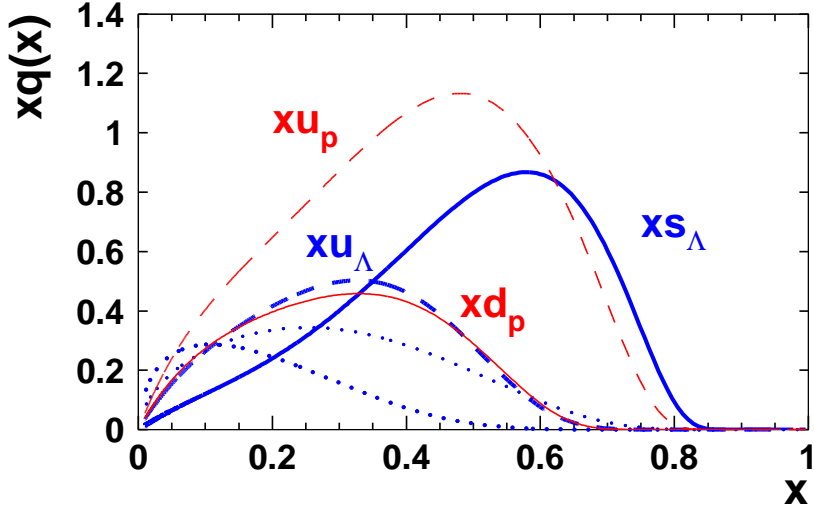,height=8.cm}
\caption{Unpolarized quark distributions in the $\Lambda$ 
(heavy lines) compared to those in the proton (light lines) 
at the bag scale, $\mu^2$. The quark distributions 
of the $\Lambda$ evolved to $Q^2=10$ GeV$^2$ are shown as 
dotted lines.} 
\label{fig1}
\end{figure}

\begin{figure}
\psfig{figure=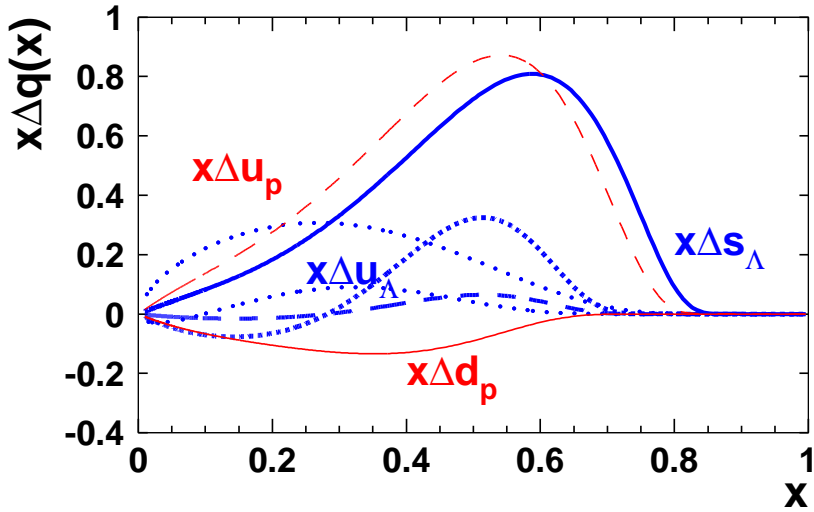,height=8.cm}
\caption{Polarized quark distributions in the $\Lambda$ 
(heavy lines) compared to those in the proton (light lines) 
at the bag scale, $\mu^2$. 
The heavy dotted line stands for five times $x\Delta u_\Lambda$
and indicates the relative importance of the $u$ and $d$
quarks in $g_1$. The quark distributions 
of the $\Lambda$ evolved to $Q^2=10$ GeV$^2$ are shown as 
dotted lines. }
\label{fig2}
\end{figure}

The calculated quark distributions  for the $\Lambda$  are shown in 
Figs. \ref{fig1} and \ref{fig2} at the scale 
relevant for the bag model, $\mu^2=0.25$ GeV$^2$,  
and with a bag radius of $0.8$ fm.   The distributions
are compared to the corresponding quark distributions in the 
proton which were also calculated in the bag model,  
using the scalar and vector diquark mass splitting fixed from  
the $\Delta$-$N$ splitting. 
We see that the quark distributions of the $\Lambda$ are quite  
different from  $SU(3)$ expectations, $s_\Lambda \ne d_p$, etc.   
Perfect $SU(6)$ symmetry would give identical up, down and strange 
distributions. The strange quark distribution is much harder than 
the up and down quark distributions.    
The polarized up and down distributions are positive for large 
$x$.  The non-strange distributions can play an important role  
whenever the strange contribution is suppressed, such as 
in processes induced by photons, where the up quark 
distributions are weighted 
by a factor of $4/9$ as opposed to $1/9$ for the strange quarks. 
Their relative magnitude is sizable as can be seen 
in Fig. \ref{fig2}, where we show five times $x\Delta u(x)$
as a dotted line to indicate the relative 
contribution of $u$ and $d$ to $g_1^\Lambda$.     

If the flavor dependence predicted for the polarization of 
non-strange quarks was retained in quark fragmentation,  
our predictions could be tested in semi-inclusive DIS  
experiments with longitudinally polarized electrons. 
Here, the smallness of the $u$ and $d$ polarizations relative
to the strange quark polarization is compensated by the
abundance of $u$-quarks in the valence region, and by the 
enhancement factor of $4/9$ for the $u$-quark (relative to the 
factor $1/9$ for $s$ quarks) in electro-magnetic
interactions \cite{Jaffe96}.  
$\Lambda$'s produced in the current fragmentation region
are mainly fragmentation products of $u$-quarks.
Part of the polarization of the electron is
transferred to the struck quark in the scattering process.
This polarization will  be transferred to the final
$\Lambda$ if the helicity dependent fragmentation
functions, $\Delta D^\Lambda_u$,  are non-zero \cite{Jaffe96}. 
In the following Section we extend our discussion to fragmentation   
and compare 
the resulting predictions with experimental data on hyperon
formation and polarization and with predictions of other models for the
fragmentation functions.  

\section{Quark fragmentation into $\Lambda$ Hyperons} 

Since the quark distributions of the $\Lambda$ baryon are not flavor 
symmetric, it is probable that the 
quark fragmentation functions are also 
flavor asymmetric. In the fragmentation 
of quarks into a specific baryon, 
quarks with different flavor couple to different 
spin-flavor components of the baryon wave function. 
For example, in order to produce  
a $\Lambda$ from an up (down) quark or a 
strange quark, the fragmentation process has to produce 
a $ds$ ($us$) vector or $ud$ scalar diquark. The mass 
differences between the scalar and vector diquarks  inevitably  
lead  to flavor dependent fragmentation 
functions, analogous to the flavor dependence of the 
quark distributions.    
 
Fragmentation functions can be defined in a manner similar to 
quark distribution functions, as light-cone Fourier 
transforms of matrix elements of quark operators   
\cite{Collins,JaffeXi} 
\begin{equation}
 \frac{1}{z} D_{q\Lambda}^{\,\Gamma}(z)
   = \frac{P^+}{2} \sum_n
   \int \frac{d\xi^-}{2\pi}
      e^{-iP^+ \xi^-/z}
         \mbox{Tr} \{ \Gamma\, \langle 0|\psi (0) |
 \Lambda (PS);n (p_n)  \rangle
        \langle \Lambda (PS); n (p_n)  
| \overline{\psi} (\xi^-) | 0\rangle \}, 
\label{frag1} 
\end{equation}
Here, $\Gamma$ is the appropriate Dirac matrix.  
Translating the matrix elements, using the integral
representation of the delta function and projecting out the 
light-cone helicity components, we obtain 
\begin{equation}
 \frac{1}{z} D^{\uparrow\downarrow}_{q\Lambda} (z)
    =  P^+ \sum_X
    \delta [(1/z-1)P^+-p_n^+] 
        | \langle 0|\psi_{+}^{\uparrow\downarrow} (0) 
|\Lambda (PS_\parallel); n (p_n)   \rangle |^2 . 
\label{frag2} 
\end{equation}
$D_{q\Lambda}^{\uparrow\downarrow}$ can be interpreted as 
the probability that 
a right-/left-handed quark fragments into a right-handed $\Lambda$ 
and similar for antiquarks. The fragmenation function of an antiquark 
into a $\Lambda$ is given by Eq. (\ref{frag2}) with  
$\psi_+$ replaced  by $\psi_+^\dagger$. 
Using Eq. (\ref{frag2}) has the 
advantage that energy-momentum    
conservation is built in {\it before} any approximation 
is made for the states in the matrix element. 
The delta function in Eq.(\ref{frag2}) implies that 
the function $D_{q\Lambda}/z$ peaks at 
\begin{equation}
   z_{max} \approx \frac{M_\Lambda}{M_\Lambda +M_n}, 
\label{max}
\end{equation}
where we have chosen to 
work in the rest frame of the $\Lambda$. 
For $M_n=2/3 M_\Lambda$ and $M_n=4/3 M_\Lambda$, we obtain
$3/5$ and $3/7$, respectively. The contributions from the  
four quark intermediate states therefore  
peak in the physical region, at relatively large
$z$ values. Thus, in contrast to the quark distribution functions, 
the fragmentation functions are not dominated by the lowest 
intermediate mass states. However, we still expect that 
for large $z$ the most important contribution comes 
from the fragmentation of a quark into a $\Lambda$ and a 
diquark state. Since the fragmentation functions are sensitive 
to the mass of the intermediate states,   
our arguments on $SU(6)$ flavor symmetry breaking  
apply in the same way to the fragmentation functions 
as to the quark distributions.  
Most importantly,  since $u^\uparrow$ and $u^\downarrow$ couple 
to different spin-flavor components of the Lambda wave function, we 
expect that not only  are the $u$ and $d$ quarks in a polarized 
$\Lambda$ hyperon  polarized, but  that  
$u$ and $d$-quarks may also fragment into a polarized $\Lambda$. 
Furthermore,  $\Delta D_{u\Lambda}$ and $\Delta D_{d\Lambda}$  
are  positive at large $z$   
for the same reason as $\Delta u$ and $\Delta d$ are 
positive at large $x$. Thus, for example,     
$\Lambda$'s produced  in the current
fragmentation region of semi-inclusive DIS processes  
should be positively polarized.  

We stress that, as for the quark distributions,  
our analysis is very general  
and follows from the definition of the fragmentation functions 
and from energy momentum conservation. 
The matrix elements can be calculated 
using model wave functions  at the scale relevant to the 
specific model and the resulting fragmentation functions can be evolved 
to a higher scale to compare them to experiments. 
First, let us  discuss the action of the operators 
in Eq. (\ref{frag2}). 
The operator,  
$\psi_+$ ($\psi_+^\dagger$ for antiquark fragmentation), 
when acting on the state $\Lambda$ in the final state:  
\begin{itemize}
\item{(i)} Can destroy a quark in $\Lambda$ leaving a diquark state 
which has to match the quantum numbers 
of the anti-diquark state  to give vacuum quantum numbers.   
This corresponds to the  fragmentation of the quark via production of two 
quark antiquark pairs from the vacuum 
$q\rightarrow (qqq) +(\bar q\bar q)$, i.e.  the fragmentation of $q$ 
into a $\Lambda$ and an anti-diquark.  
\item{(ii)} Alternatively, $\psi_+$  
($\psi_+^\dagger$) can insert  an antiquark (quark) into the 
$\Lambda$ wave function.  
In this case, the intermediate state must  be a four-quark state such that
vacuum quantum numbers are preserved.  
This corresponds to the fragmentation of a  quark  (antiquark)   
into a $\Lambda$  via production of three quark antiquark pairs,  
$q \rightarrow  (qqq) + (q \bar q\bar q\bar q)$ 
($\bar q \rightarrow  (qqq) + (\bar q \bar q\bar q\bar q)$). 
\end{itemize}

In order to quantify our discussion we have to use 
model wave functions for the states. 
Choosing the Peierls-Yoccoz projection 
method and MIT bag wave functions,  the fragmentation functions 
are given by 
\begin{equation}
D_{f\Lambda}^{\uparrow\downarrow}(x) =
\frac{z M_\Lambda}{(2\pi )^2} \sum_{m} 
\langle \Lambda^\uparrow |P_{f,m}|\Lambda^\uparrow   \rangle
\int^\infty_{p_{min}}
p_ndp_n |\langle 0|\psi_+^{\uparrow\downarrow} (0) 
|\Lambda (PS_\parallel ); n (p_n)   
 \rangle |^2.   
\label{eq:frag}
\end{equation} 
with 
\begin{equation}
   p_{min}=|\frac{M_\Lambda^2(1-z)^2-z^2M_n^2}
 {2M_\Lambda z(1-z)}|. 
\end{equation}
The matrix element for the fragmentation through a 
diquark intermediate state yields 
\begin{equation} 
\langle 0|\psi_+^{\uparrow\downarrow} (0) 
|\Lambda (PS_\parallel ); n (p_n)   \rangle  = 
 [\phi_2 ({\bf p_n}) \phi_3({\bf 0})]^{-1}
   \hat{\psi}_+^{\uparrow\downarrow} (-{\bf p_n})
\int d{\bf R} \,\, e^{i{\bf p_nR}}
 \langle 0
   | B_2({\bf 0}) B_{\bar 2 }({\bf R}) \rangle  .
\end{equation} 
Here, $\phi_2({\bf p_n})$ and $\phi_3(0)$ are normalization constants 
of the final states. 
$\hat{\psi}_+^{\uparrow\downarrow}(-{\bf p_n})$ is the Fourier transform 
of the projected  bag wave function. 
The matrix element, 
$\langle 0 | B_2({\bf 0}) B_{\bar 2 }({\bf R}) \rangle$,   
describes   the transition between  the diquark anti-diquark states 
and the vacuum.   
We assume that it is proportional to the overlap of the diquark and 
anti-diquark states with a $\gamma_0$ sandwiched between them    
$\langle \bar B_{\bar 2 }({\bf R})| B_2({\bf 0})\rangle$.    
\footnote{This factor is basically  a normalization factor. 
In the case of the quark distributions, the corresponding 
expression gives just $\phi_2({\bf p_n})$. 
It has a  small effect on the shape of the fragmentation 
functions which is mainly determined by the kinematic 
constraints and the Fourier transform of the wave  function 
of the struck quark.}  
We will calculate this overlap and adjust 
the  normalization constant by fitting one data point later 
when we discuss the phenomenological implications 
of our fragmentation functions.      
The expression for the four quark intermediate states 
can be obtained by replacing $\phi_2$, $B_2$ and $B_{\bar 2}$ 
through $\phi_4$, $B_4$ and $ B_{\bar 4}$      
and by replacing the positive 
energy, ground state bag solutions,  
$\hat{\psi}_{+1s}^{\uparrow\downarrow}({\bf -p_n})$       
by $\hat{\psi}^{\dagger\uparrow\downarrow}_{+1s}({\bf p_n})$ or by 
the corresponding negative energy state 
$\hat{\psi}_{+1\bar s}^{\uparrow\downarrow}({\bf -p_n})$ 
for the three anti-quark one quark intermediate state or the    
four anti-quark intermediate state,  respectively \cite{Schreiber}.      

Denoting by $\hat{F}(z)$ and $\hat{G}(z)$ those contributions to Eq.
(\ref{eq:frag}) which come from the $f( {\bf - p_n})$ and
$g({\bf - p_n})$ parts of the integral,  
the contributions from the diquark  intermediate states to the 
fragmentation functions are then given by 
\begin{eqnarray} 
   D_{d\Lambda} (z) & = & 
 D_{u\Lambda} (z) = \frac{1}{4}[3\widehat{F}_v(z)+\widehat{F}_s(z)]
\nonumber \\ 
 D_{s\Lambda} (z) & =&  \widehat{F}_s(z) \nonumber\\ 
\Delta D_{d\Lambda} (z) & = & \Delta D_{u\Lambda} (z)  = 
 \frac{1}{4}[\widehat{G}_s(z) - \widehat{G}_v(z)] \nonumber \\ 
\Delta  D_{s\Lambda} (z) & = & \widehat{G}_s(z),  
\end{eqnarray} 
with $\Delta D_{q\Lambda} \equiv D^\uparrow_{q\Lambda} 
- D^\downarrow_{q\Lambda}$.  
Since $\widehat{G}_s$ is shifted towards positive 
$z$ values relative to $\widehat{G}_v$,   
$\Delta D_{d\Lambda}$ and $\Delta D_{u\Lambda}$ 
are positive for large $z$. 
The results are shown in Figs. \ref{fig3} and  \ref{fig4}. 
We show the results both  at the bag scale $\mu^2 = 0.25$ GeV$^2$ and 
at  $Q^2=10$ GeV$^2$. We evolved the fragmentation functions 
in leading order  and set the gluon fragmentation function 
to zero at the starting scale for the non-singlet evolution. 
We used the package of Ref. \cite{Kumano} modified for 
the evolution of fragmentation functions.

The fragmentation functions possess the following qualitative 
features:  
\begin{itemize}
\item{(1)} the mass splitting associated with the hyperfine interaction 
leads to considerable $SU(6)$ breaking in the fragmentation functions; 
\item{(2)} at large $z$, fragmentation functions are dominated by 
the diquark intermediate states; 
\item{(3)} since the contributions of higher mass states 
to fragmentation functions have maxima 
in the physical region, they play an important role at lower 
$z$ values; 
\item{(4)} the splitting of the vector and scalar diquark masses 
leads to polarized, non-strange fragmentation functions.  
\end{itemize}

\section{Phenomenology} 

\subsection{$e^+e^-$ annihilation at the $Z$ resonance}

Before making predictions for the expected $\Lambda$ 
polarization in semi-inclusive DIS, we check whether available  
$\Lambda$ production data are consistent with 
our approach. 

\begin{figure}
\psfig{figure=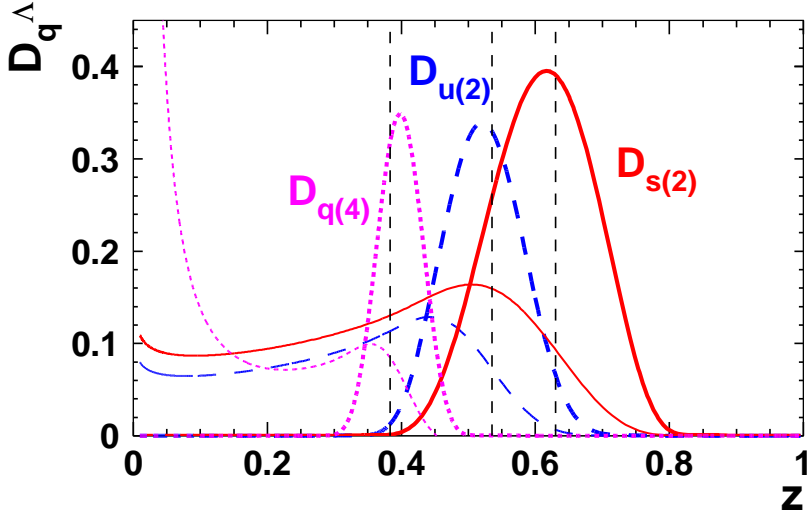,height=8.cm}
\caption{Contributions of two and four quark intermediate 
states to the unpolarized fragmentation functions at the bag scale 
(heavy lines) and evolved to $Q^2=10$ GeV$^2$ (light lines). 
The mass of the four-quark intermediate state is set to $1.8$ GeV and 
$D_{q(4)}$ is a sum over flavors,  \protect $D_{q(4)} = 
D_{\bar u\Lambda}+ D_{\bar d\Lambda}+D_{\bar s\Lambda}$.  
The dashed vertical lines indicate the position of the maxima 
using Eq. (\protect\ref{max}). }
\label{fig3}
\end{figure}

\begin{figure}
\psfig{figure=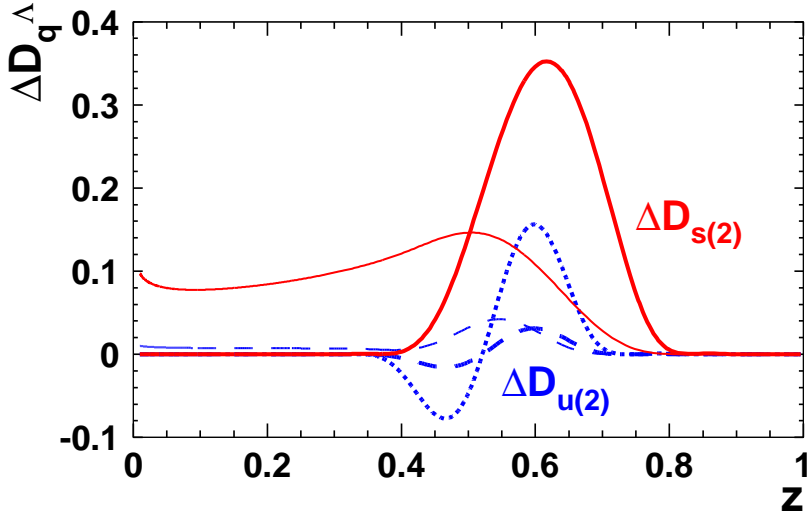,height=8.cm}
\caption{Contributions of two and four quark intermediate  
to the polarized fragmentation functions at the bag scale 
(heavy lines) and evolved to $Q^2=10$ GeV$^2$ 
(light lines).  
The dotted line stands for five times $\Delta D_{u}^\Lambda$. }
\label{fig4}
\end{figure}

Let us start our discussion with an interesting consequence of our 
approach for the lower-$z$ end of the spectrum of  particles 
produced in $e^+e^-$ annihilation. When the number of particles 
produced  is plotted as a function of $\xi=\ln(1/z)$, one finds 
that the spectra  
exhibit an  approximate Gaussian shape around a maximum, $\xi^*$, which  
depends on the produced particle \cite{Delphixi,L3xi,Opalxi,SLD}.  
While the shape of the spectrum can be understood 
in perturbative QCD as a consequence of the coherence 
of gluon radiation \cite{Dok}, the position of the maximum is 
a free parameter which has to be extracted from  experiment.  
Energy-momentum conservation dictates 
that the spectrum  at small $z$ is dominated 
by high mass intermediate states. At a given total 
invariant mass $\sqrt{s}$, there will
be a maximum value for the mass of the intermediate state which can be
produced in the fragmentation. This maximal mass determines
the ``lower'' edge of the spectrum. Note also, that because of the 
$1/z$ factor in the lower limit of the integration in Eq. 
(\ref{eq:frag}), the fragmentation function drops at low $z$ 
values.  The $\xi$-distribution 
is given by  $d\sigma/d\xi= zd\sigma/dz 
\sim z D(z)$, and thus it is proportional to $z^2$ times 
$D(z)/z$. Although Eq.(\ref{max}) describes the location of 
the maxima of the distribution $D(z)/z$,  
we can expect that Eq.(\ref{max}) is also a good approximation for 
the $\xi$-distribution since the fragmentation functions for 
a given mass $M_n$ are very narrow, as can be seen in Fig. \ref{fig3}. 

\begin{figure}
\psfig{figure=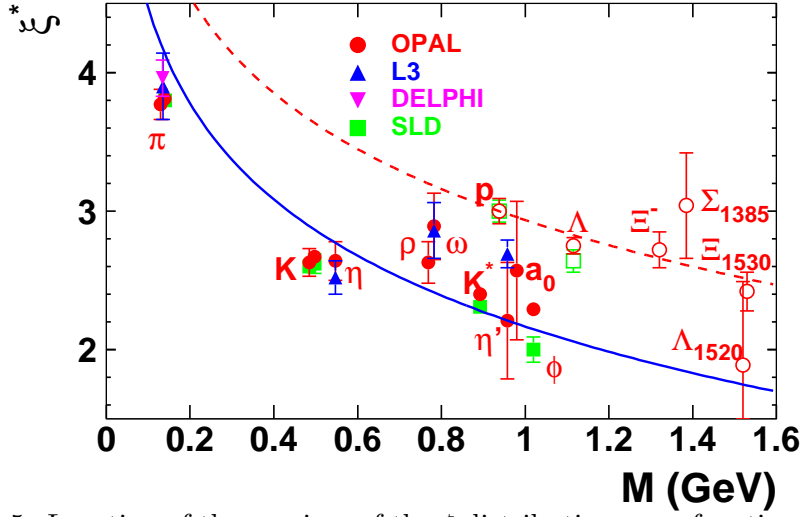,height=8.cm}
\caption{Location of the maxima of the $\xi$ distributions 
as a functions of the particle mass. The full and open symbols 
represent mesons and baryons, respectively. The data are from 
\protect\cite{Delphixi,L3xi,Opalxi,SLD}. The solid  
and dashed lines 
are the prediction of Eq. (\protect\ref{maxxi}) 
for mesons and baryons adjusting the 
normalization to $\xi^*_{\eta^\prime}$ 
and to  $\xi^*_p$, respectively.}  
\label{fig5}
\end{figure}

From Fig. \ref{fig3}, we can also see that Eq. (\ref{max}) is a good 
approximation for the maximum of $D(z)$. 
(Eq. (\ref{max}) gives the maximum of the $\xi$-distribution  
exactly in the limit when the distribution is a $\delta$-function.)  
In the following, we will use Eq. (\ref{max}) to estimate the maxima 
of the $\xi$-distribution. Now, the maximum of the
distribution coming from this highest mass state determines the
maximum of the fragmentation function in first approximation
through Eq. (\ref{max}). Although  $M_n$ is not known, it 
should be proportional to  
$\sqrt{s}$. Thus, the maximum of the $\xi$ distribution has  
the correct  $\ln(s)$ dependence as seen in the 
experiment \cite{Delphixi,L3xi,Opalxi,SLD}.    
However, if we take the difference 
of the maxima of the $\xi=\ln(1/z)$ distributions of different 
particles, this dependence on the unknown maximum value of $M_n$  
drops out for large $M_n$. It follows from Eq.(\ref{max}) that 
\begin{equation}
  \Delta \xi^* = \xi^*_a -\xi^*_b 
  \approx \ln \left( \frac{M_a + M_n}{M_b + M_n}\right) +
  \ln \frac{M_b}{M_a} \approx \ln \frac{M_b}{M_a} ~.
\label{maxxi}  
\end{equation}
Thus, the difference of the maxima is determined by 
the logarithm of the ratio 
of the masses. 
We calculated the maxima of the $\xi$ distributions
using this formula and taking the maximum of the $\eta^\prime$ 
and the proton distributions as a reference value 
for mesons and baryons, respectively. 
The results are compared 
to the experimental data \cite{Delphixi,L3xi,Opalxi,SLD}   
in Fig. \ref{fig5}.  
We stress that our results follow from the general definition 
of the fragmentation functions and from energy-momentum conservation 
and are in remarkably good agreement with the data.

Let us turn our attention to the high $z$ region 
where, according to our discussion in Sect. II,   
significant flavor symmetry breaking effects are to be 
expected. There are experimental data for the production of 
$\Lambda$ hyperons in $e^+e^-$ annihilation. 
A considerable fraction of the data was taken at the Z-resonance,  
where the quarks produced in the annihilation process 
are longitudinally polarized.  The spin dependent 
fragmentation functions can then be determined by 
measuring the polarization of the produced Lambda baryons.  

In the quark parton model, 
the cross section for the inclusive production of 
a Lambda hyperon, $e^+e^- \rightarrow \Lambda +X$,  
is obtained by summing over the cross sections for 
$e^+e^- \rightarrow q\bar q$ and weighting with the  
probabilities $D_{q\Lambda}$ that a quark fragments 
into a Lambda with  energy fraction $z$,  
\begin{equation} 
   \frac{d^2\sigma_\Lambda}{dz d\Omega} = 
 \sum_q    \frac{d\sigma^q}{d\Omega} \left[ D_{q\Lambda} (z,Q^2) + 
       D_{\bar q\Lambda} (z,Q^2) \right].  
\end{equation} 
Here, z is defined as $z=2p_\Lambda\cdot q/Q^2= 2 E_\Lambda/\sqrt{s}$ 
for c.m.s energy $\sqrt{s}$, and $p_\Lambda$ and $E_\Lambda$ 
are the four-momentum and energy of the $\Lambda$;  
$q$ and $Q^2\equiv q^2=s$ are respectively the four-momentum and 
invariant mass of the virtual $Z$-boson.

After integrating over angles, the cross section 
at the Z-resonance can be written as 
\begin{equation} 
  \frac{d\sigma_\Lambda}{dz} = \frac{4\pi \alpha^2}{s} 
    \sum_q \hat{e}^2_q  
    \left[ D_{q\Lambda} (z,Q^2) +
       D_{\bar q\Lambda} (z,Q^2) \right], 
\label{cross} 
\end{equation}  
where 
the coupling of the quarks to the $Z$ is given by 
\begin{eqnarray} 
  \hat{e}^2_q  &=&   e_q^2 +   (1+v_e^2)(1+v_q^2) \chi (M_Z^2) 
  \nonumber \\ {\rm with}~~~ 
  \chi (s)  &=&  \frac{1}{256 \sin^4\Theta_W \cos^4\Theta_W} 
     \frac{s^2}{(s-M_Z^2)^2 + \Gamma_Z^2 M_Z^2}~~,
\end{eqnarray} 
where $M_Z$ and $\Gamma_Z$ are the mass and width of the $Z$. 
$v_e=-1+4\sin^2\Theta_W$, $v_u=1-\frac{8}{3} \sin^2\Theta_W$, 
$v_d=-1+\frac{4}{3}\sin^2\Theta_W$   
are the  vector coupling 
of the electron and the quarks to the $Z$    
\footnote{Since the data are taken at the Z-resonance peak, 
we have dropped terms in Eq.\ (\protect\ref{cross}) which cancel for 
$s=M_Z^2$. Note also that the terms proportional to $e^2_q$ are very 
small and can be neglected in numerical calculations.}.  

In the following, we introduce   ``valence'' and ``sea'' type  
fragmentation functions 
$D_{q_v\Lambda} \equiv D_{q\Lambda} - D_{\bar{q}\Lambda}$, 
and $D_{q_s\Lambda}\equiv D_{\bar q\Lambda}$.  
Eq. (\ref{cross}) can then 
be re-written as  
\begin{equation} 
  \frac{d\sigma_\Lambda}{dz} = \frac{4\pi \alpha^2}{s} 
    \sum_q \hat{e}^2_q  
    \left[ D_{q_v \Lambda} (z,Q^2) +
      2 D_{q_s \Lambda} (z,Q^2) \right], 
\label{cross2} 
\end{equation}  

Experimental measurements show that 
produced particles containing the initial quark  as a valence 
quark have higher momenta than the other hadrons in the jet --- 
this is the ``leading particle'' effect. For example, $\Lambda$'s 
produced in 
a light quark jet have higher momenta than $\bar\Lambda$,  
indicating that the fragmentation functions $D_{q\Lambda}$ 
are harder than $D_{q\bar\Lambda}$, or by CP invariance 
harder than $D_{\bar q\Lambda}$. Thus,   
the flavor non-singlet combination of the 
fragmentation functions,  
$D_{q_v\Lambda} \equiv D_{q\Lambda} - D_{\bar{q}\Lambda}$, 
effectively measures leading particle production and 
can be identified with the contribution from 
diquark component of the fragmentation functions, for obvious reasons.  
On the other hand, particles produced in the  
fragmentation process 
populate more the central rapidity region and are 
likely to be independent of the flavor of the 
initial quarks.   
Thus, the ``sea'' type fragmentation functions 
$D_{q_s\Lambda}\equiv D_{\bar q\Lambda}$  
can be identified with the contributions from intermediate four quark 
higher mass states.

\begin{figure}
\psfig{figure=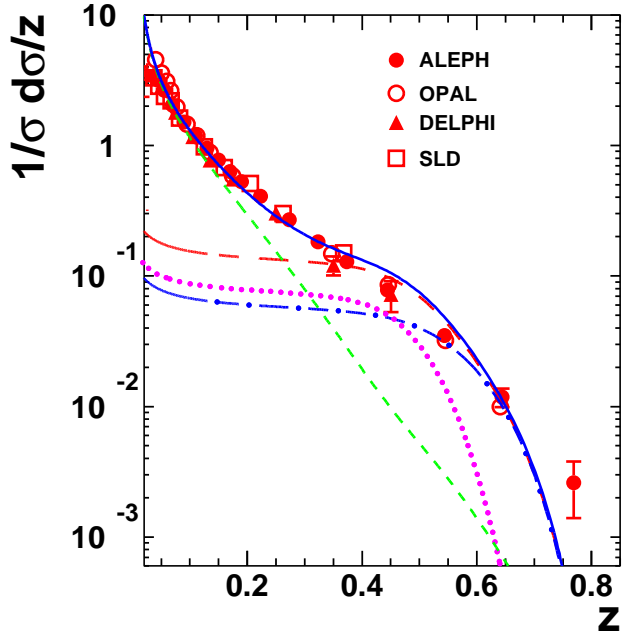,height=10.cm}
\caption{The inclusive cross section  
$\frac{1}{\sigma_{tot}} \frac{d\sigma^\Lambda}{dz}$  
in $e^+e^-$ annihilation at the $Z$ resonance. 
The dash-dotted and the short dashed lines represent 
the contributions from $D_{s_v}$ and $D_{u_v}+D_{d_v}$, 
respectively; the dashed line is the total $D_{q_v}$ 
contribution and the dotted line stands for $D_{q_s}$. 
The data are from Refs. \protect\cite{Aleph,Delphi,L3,Opal,SLD}.}
\label{fig6}
\end{figure}

In the following, we use our calculated fragmentation functions 
for the valence contribution and parameterize the sea  part,  
since the mass of the higher mass four quark states is not 
well-known.  
As the production of non leading particles is independent 
of the flavor of the initial quarks, it 
is reasonable to assume that the sea type fragmentation functions are 
flavor symmetric --- i.e. $D_{q_s\Lambda}\equiv D_{u_s\Lambda} = 
D_{\bar u\Lambda} = D_{d_s\Lambda} = .... =D_{\bar b\Lambda}$.   
As the $e^+e^-$ experiments measure the sum of $\Lambda$ 
and $\bar\Lambda$ production \cite{SLD,Aleph,Delphi,L3,Opal}, 
we use CP invariance to relate the fragmentation 
functions of the $\Lambda$ to those of the $\bar\Lambda$, i.e.  
$D_{q\Lambda} =D_{\bar q\bar\Lambda}$ and
$D_{\bar q\Lambda} =D_{q\bar\Lambda}$. 
We evolve the fragmentation functions   
in leading order $Q^2$-evolution from the scale relevant to 
the bag model, $\mu^2=0.25$ GeV$^2$, to $M_Z^2$ in order to 
obtain $D_{q_v\Lambda}(z,Q^2=M_Z^2)$ .  
In order to evolve the singlet part of the fragmentation functions,  
we assume that the gluon fragmentation functions are zero 
at the starting scale. 
Our result is shown in Fig. \ref{fig6} in comparison with  
data from SLD  \cite{SLD} and  
the LEP experiments \cite{Aleph,Delphi,L3,Opal}. 
The long dashed line represents the contribution of $D_{q_v}$, 
the dotted line is the fitted sea quark fragmentation $D_{q_s}$ 
and the solid
line is the sum of the two contributions. The dash-dotted and 
short dashed lines are the contributions of strange  
and up plus down valence terms, respectively. 
We do not attempt to reproduce the data at low $z$ values 
where gluon coherence effects become important and 
the usual  evolution equations break down \cite{Dok}.    
The data clearly favors a two component picture.

\begin{figure}
\psfig{figure=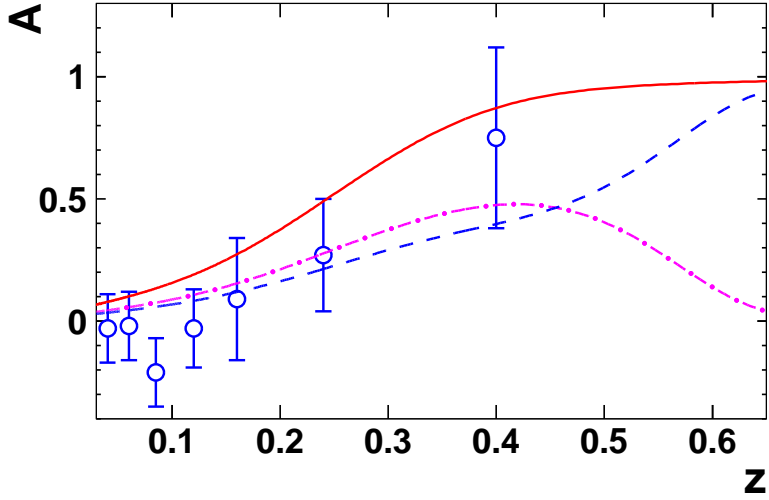,height=8.cm}
\caption{The asymmetry between leading and non-leading 
Lambda production. The solid line 
is the prediction of the two component picture. 
The dashed and dash-dotted lines are the contributions 
from the strange and from the up plus down quarks, respectively. 
The data are taken from Ref. \protect\cite{SLD}.} 
\label{fig7}
\end{figure}

The asymmetry in  
leading and non leading particle production in $e^+e^-$ annihilation 
provides a further test for the fragmentation functions. 
This asymmetry has been measured by the SLD Collaboration \cite{SLD}  
and is defined as   
\begin{equation} 
   A=\frac{R_\Lambda^q - R^q_{\bar\Lambda}} 
        {R_\Lambda^q + R^q_{\bar\Lambda}} 
\end{equation} 
with 
\begin{eqnarray} 
   R_\Lambda^q & = & \frac{1}{2N_{ev}} 
          \frac{d}{dz} \left[ N(q\rightarrow \Lambda ) 
               + N(\bar q \rightarrow \bar\Lambda ) \right] \nonumber \\ 
   R_{\bar\Lambda}^q & =& \frac{1}{2N_{ev}} 
          \frac{d}{dz}\left[ N(q\rightarrow \bar \Lambda ) 
               + N(\bar q \rightarrow \Lambda ) \right]
\end{eqnarray} 
where $dN[q(\bar q)\rightarrow \Lambda (\bar\Lambda )]/dz$ 
is the number density of $\Lambda$'s  ($\bar\Lambda$'s) 
produced in a $q$ ($\bar q$) jet normalized to the total number 
of events $N_{ev}$. While a zero value of the asymmetry corresponds to 
equal production of hadrons and anti-hadrons,  
a value of $+1$ ($-1$) corresponds to total 
dominance of hadron (anti-hadron) production.  
In our model, the asymmetry is given by 
\begin{equation} 
    A= \frac{ \sum_q \hat{e_q}^2 D_{q_v\Lambda} (z,Q^2) } 
       {\sum_q \hat{e_q}^2 \left[ D_{q_v\Lambda} (z,Q^2)  
         + 2 D_{q_s\Lambda} (z,Q^2)\right]} . 
\end{equation} 
Here, the summation runs only over the light flavors 
since only light quark jets were used in the experimental 
determination of the asymmetry.  
Our results are compared to the data in Fig. \ref{fig7}. 
The dashed and dash-dotted lines are the strange quark and 
up plus down quark contributions, and the solid line is their sum. 
Our model gives reasonable agreement with $A$ as measured by SLD.  
At high $z$, the initial strange quarks give the dominant leading 
particle contribution to the asymmetry.  

Unpolarized $e^+e^-$ cross section measurements clearly 
support a two component picture of quark fragmentation. 
However, they cannot differentiate between a flavor symmetric model 
and a flavor asymmetric picture for the fragmentation 
functions.  For example, both the total 
cross section and the asymmetry between leading and non-leading 
particle production could be described in a model 
with flavor symmetric fragmentation functions.  Therefore, 
more information is needed to differentiate between the two
models.  Polarization measurements at the $Z$ resonance 
can give additional constraints on the fragmentation functions. 

The initial quarks produced in $e^+e^-$ annihilation at the $Z$ resonance 
are polarized because the parity violating coupling 
of the fermions favors certain helicity states. This initial 
polarization of the produced quarks can be 
transferred to the final state hadrons and may lead to polarized 
Lambda production \cite{Jaffe93}.     
The difference between the cross sections to produce 
left or right-handed Lambdas at the Z-resonance is given by 
\cite{Jaffe93}  
\begin{equation}
  \frac{d\Delta\sigma}{dz} \equiv  
  \frac{d\sigma_L}{dz}- 
   \frac{d\sigma_R}{dz} 
   = \frac{4\pi \alpha^2}{s} \sum_q \hat{g}_q 
\left[ \Delta D_{q\Lambda} (z,Q^2) 
 -\Delta D_{\bar q\Lambda} (z,Q^2)  
        \right] , 
\label{diff}
\end{equation}
with the effective couplings 
\begin{equation}
  \hat{g}_q =     2 a_q v_q (1+v_e^2) \chi(M_Z^2). 
\end{equation}
Here,  $a_u=-a_d=1$  
are the  axial vector coupling of the quarks to the Z  
and $\chi$ was defined in Eq. (20). 
We  introduce valence and sea type polarized 
fragmentation functions 
$\Delta D_{q_v\Lambda} \equiv \Delta D_{q\Lambda} - 
\Delta D_{\bar{q}\Lambda}$ and  
$\Delta D_{q_s\Lambda}\equiv \Delta D_{\bar q\Lambda}$ similar to 
the unpolarized case.  According to our interpretation, 
the valence type fragmentation functions are given by  
the fragmentation of  quarks into a $\Lambda$ and an anti-diquark. 

Since the hyperfine interaction leads to   
polarized up and down quark distributions for the Lambda,  
we expect that 
polarized up and down quarks may also fragment into 
polarized $\Lambda$'s. 
On the other hand, $\Delta D_{q_s\Lambda}$ 
is given by the contributions of four quark states. 
Since these are insensitive to the $SU(6)$  wave function of the 
$\Lambda$ we expect that they do not contribute to polarized 
Lambda production. 
Note that since the difference of the 
cross sections in Eq. (\ref{diff}) is proportional only to the 
valence part of the polarized fragmentation functions, the former 
assumption is not necessary for the description of the 
cross section difference. 

\begin{figure}
\psfig{figure=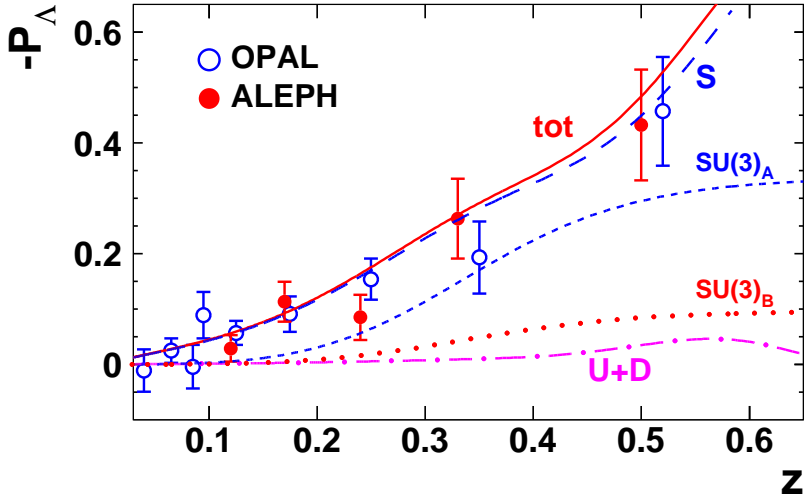,height=8.cm}
\caption{Lambda polarization in $e^+e^-$ annihilation 
at the $Z$-resonance. The solid line 
is the prediction of the two component picture. 
The dashed and dash-dotted lines are the contributions 
from the strange and from the up plus down quarks.   
The short dashed and dotted lines correspond to the 
prediction of a model with flavor symmetric 
fragmentation functions using the naive quark parton model 
($SU(3)_A$), and $g_1$ measurements 
for the proton plus $SU(3)$ flavor symmetry to relate 
the polarized fragmentation functions to  
the unpolarized ones ($SU(3)_B$). 
The data are taken from Ref. \protect\cite{Alephp,Opalp}.} 
\label{fig8}
\end{figure}

The calculated Lambda polarization, 
$P_\Lambda = - \Delta\sigma/\sigma$, is compared to the 
Aleph \cite{Alephp} and OPAL data \cite{Opalp} in Fig.\ \ref{fig8}. 
The dominant contribution comes from the fragmentation of 
strange quarks (the dash-dotted curve in Fig.\ \ref{fig7}). The 
contribution from up and down quarks (dashed curve) is negligible 
because the corresponding polarized valence type fragmentation 
functions are small.  The nonstrange quark contribution to the
$\Lambda$ polarization also peaks at lower $z$ values where 
non-leading particle production dominates. In the limit 
$z\rightarrow 1$, both $\Delta\sigma$ and $\sigma$ are dominated by 
strange quark fragmentation and $P_\Lambda$ behaves like 
\begin{equation} 
 \lim_{z\rightarrow 1}  P_\Lambda (z) 
  = \frac{2 v_d}{(1+v_d^2)} \frac{\Delta D_{s_v\Lambda}}
       {D_{s_v\Lambda}} \approx - 0.94 \,\, \frac{\Delta D_{s_v}^\Lambda}
   {D_{s_v\Lambda}} ~~, 
   \label{pcalc}
\end{equation}
where we have used $\sin^2\theta_W \approx 0.23$.  Since 
$\Delta D_{s_v\Lambda} \rightarrow D_{s_v\Lambda}$ in the limit 
$z\rightarrow 1$, 
$P_\Lambda(z)$ approaches $-0.94$ for $z\rightarrow 1$.    
In deriving Eq.\ (\ref{pcalc}), we neglected the production of higher mass 
hyperons and their subsequent decays into $\Lambda$. If these hyperons 
are produced by the initial quarks they are, in general, also polarized 
and can transfer part of their polarization to the final 
$\Lambda$. However, the distribution of the $\Lambda$'s 
coming from the decays of such hyperons should be shifted towards 
smaller $z$ values, where the contribution from 
the $q_{s\Lambda}$ is large and this will strongly suppress   
any polarization. \footnote{For a discussion 
of the effect of hyperon decays on the $\Lambda$ polarization see 
Refs. \protect \cite{Gustavson,Boros98}.}   

We can contrast our predictions with those from models with flavor 
symmetric fragmentation functions. 
We set $D_{s\Lambda}=D_{u\Lambda}=D_{d\Lambda}$ 
and discuss two different scenarios 
proposed in the literature \cite{Jaffe93,deFlorian98a,deFlorian98b}  
for the spin dependent fragmentation functions \footnote{See 
Ref. \protect\cite{deFlorian98b} for a next to leading order analysis 
along the same lines.}:  

(A) Naive quark model inspired scenario: all spin dependent 
fragmentation functions except for $\Delta D_{s\Lambda}$ are 
zero. 

(B) Predictions based on $SU(3)$ flavor symmetry and 
polarized DIS on nucleon targets: not only 
the $\Delta D_{s\Lambda}$ but also $\Delta D_{u\Lambda}$ and 
$\Delta D_{d\Lambda}$ are non-vanishing. 

In both cases, we set the polarized valence type fragmentation 
functions proportional to the unpolarized ones. 
\begin{equation}
   \Delta D_{s\Lambda} (z) = c_s  D_{s\Lambda}(z);  
  \,\,\,\,\,\,\,\,    \Delta D_{u\Lambda} (z)
=    \Delta D_{d\Lambda} (z)  = c_u  D_{u\Lambda}(z),
\label{relations} 
\end{equation} 
In case (A), we have $c_s=1$ and $c_u=0$ and 
in case (B), we have $c_s=0.6$ and $c_u=-0.2$. 
We could allow a different $z$-dependence between the 
polarized and unpolarized quark distributions by multiplying 
by a power of $z$, for example. However, we are interested 
in the upper limits one can obtain from flavor symmetric models.   
Since the polarized fragmentation functions are 
bounded by the positivity constraint, $|\Delta D_q|\le D_q$, 
we have not included a suppression factor.  

The limiting behavior of the polarization is 
\begin{eqnarray} 
 \lim_{z\rightarrow 1}  P_\Lambda^{(A)} (z) 
  &=& \frac{2 v_d}{(1+v_u^2)+2(1+v_d^2)}
   \approx - 0.34 \nonumber \\ 
 \lim_{z\rightarrow 1}  P_\Lambda^{(B)} (z)
  &=& \frac{2[(c_u+c_s)v_d - c_u v_u]}{(1+v_u^2)+2(1+v_d^2)} 
  \approx -0.1 ,
\end{eqnarray}
for case (A) and (B), respectively. 
The unpolarized cross sections were fit using 
flavor symmetric fragmentation functions for both 
the valence and sea type fragmentation functions, and 
the Lambda polarization was calculated using Eq.\ (\ref{relations}). 
The results are shown in Fig. \ref{fig8}. 
Even when the positivity constraint is saturated, 
{\it i.e.} $\Delta D_{s\Lambda} = D_{s\Lambda}$, the  
Lambda polarization for high $z$ values is considerably smaller 
than experiment in both cases.  
The data strongly suggest that the 
fragmentation functions cannot be flavor symmetric.  

In this connection, we note that  
Monte Carlo models often used by the experimental 
Collaborations, {\it e.g.} JETSET \cite{JETSET}  
which is based on the Lund fragmentation model \cite{Lund}, 
can describe the $P_\Lambda$ data using 
parameters obtained from the naive quark model for the 
$\Lambda$ spin content. 
However, these Monte Carlo programs have built in 
parameters which suppress 
\begin{itemize} 
\item {(a)} the production of strange 
quarks relative to up and down quarks, 
\item{(b)} the production of 
strange diquarks relative to diquarks containing only up and down quarks, 
and 
\item{(c)}  the production of 
vector diquarks relative to scalar diquarks. 
\end{itemize} 
These suppression factors result in {\it flavor asymmetric} 
fragmentation functions. For example, an initial strange quark 
has a higher probability to fragment into a Lambda than 
an up or down quark, due to the suppression for the production of 
strange diquarks from the vacuum. 
The suppression factor (c) makes it straightforward 
to implement our ideas in a polarized version of such Monte 
Carlo programs. 

\subsection{Semi-inclusive deep-inelastic scattering} 

We have seen that, in $e^+e^-$ annihilation, 
a flavor separation of the polarized fragmentation 
function is not possible because the Lambda polarization 
is dominated by the fragmentation of strange   
quarks.  However, since fragmentation of up quarks is the dominant  
channel for Lambda production in semi-inclusive DIS,  
this process is very useful for studying the polarized 
up quark distribution functions, as was pointed out 
in Ref. \cite{Jaffe96}. 

\begin{figure}
\psfig{figure=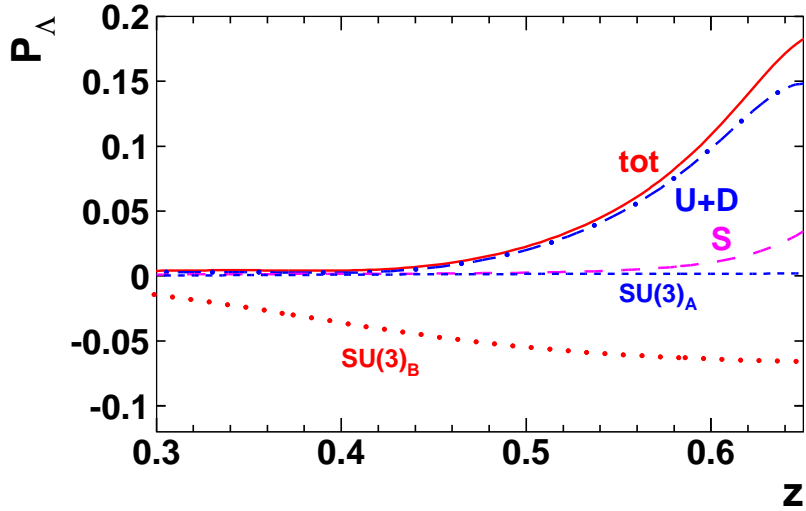,height=8.cm}
\caption{The polarization of the $\Lambda$ produced in semi-inclusive,
polarized $\vec{e}-p$ scattering. The results were calculated for
$E_e = 30$ GeV, $x=0.3$, and $Q^2 = 10$ GeV$^2$. The electron 
polarization is arbitrarily set to 50\%. The contributions from
the fragmentation of $u+d$ and $s$ quarks are shown
as dash-dotted and dashed lines, respectively. The solid line is the
total polarization. The predictions of the flavor symmetric models
(A) and (B) are shown as short dashed and dotted lines, respectively.}
\label{fig9}
\end{figure}

The Lambda polarization resulting from the scattering of polarized 
electrons from an unpolarized nucleon target is given by \cite{Jaffe96}
\begin{equation}
  \vec{P}_\Lambda = \hat{e}_3  P_e \frac{y(2-y)}{1+(1-y)^2}
 \frac{\sum_q e_q^2 q_N(x,Q^2) \,\Delta D_{q\Lambda} (z,Q^2) }
{\sum_q e^2_q q_N(x,Q^2)\, D_{q\Lambda} (z,Q^2)},
\end{equation}
where $y\equiv (E-E^\prime)/E$ is the usual DIS variable and 
$z\equiv p_\Lambda \cdot p_N/p_N \cdot p$ where  
$p_\Lambda$, $p_N$ and $q$ are the four-momenta 
of the $\Lambda$, nucleon and the virtual photon.   
The electron beam defines the $\hat{e}_3$ axis and
$P_e$ is the degree of polarization of the incident electron. 
At not too small Bjorken $x$-values the 
contributions from strange quarks may be neglected,  
and $P_\Lambda$ measures effectively  
$\Delta D_{u\Lambda} /D_{u\Lambda}$.  
We calculated the $\Lambda$ polarization
using  our  fragmentation functions.  
Fig. \ref{fig9} shows the 
result calculated at $E_e\approx 30$ GeV, $x=0.3$ and $Q^2=10$ GeV$^2$, 
where $y=0.58$.  A beam polarization of $50\%$ was assumed.
The dash-dotted and dashed lines are the contributions from the 
fragmentation of $u$ plus $d$ quarks and $s$ quarks, respectively. The 
solid line is the total polarization. We see that the polarization is 
positive and large for higher $z$ values, and that 
the dominant contributions come from the fragmentation of up quarks. 
At even larger $z$ values, the contribution of strange quarks 
becomes important since $\Delta D_{s\Lambda}$ is harder than  
$\Delta D_{u\Lambda}$, as can be seen from Fig.\ \ref{fig2}.  
However, since the
cross section decreases rapidly with increasing $z$, the bulk of the
produced $\Lambda$'s are fragmentation products of
$u$-quarks.  Thus, in semi-inclusive scattering of polarized 
electrons from nucleons, a positive value of $P_\Lambda$ at 
intermediate values of $z$ would confirm our prediction. 
Although the absolute values of $\Delta D_{u_v\Lambda}$,  
are quite small, they lead to a relatively large polarization 
since, in the limit $x \rightarrow 1$ ($z \rightarrow 1$),   
the component of the wave function containing a scalar diquark 
dominates both  $\Delta D_{u_v\Lambda}$ and 
$D_{u_v\Lambda}$.  In Fig. \ref{fig9}, we also show the 
predictions resulting from flavor symmetric fragmentation functions 
for both cases (A) and (B). While (A) gives essentially 
zero $\Lambda$ polarization, (B) predicts a negative 
$\Lambda$ polarization due to the fragmentation of 
negatively polarized up quarks.  Thus, polarization measurements in 
semi-inclusive DIS can easily differentiate between our    
predictions and those obtained from flavor symmetric models.   

\section{Conclusions} 

We discussed the quark parton structure of the $\Lambda$ 
baryon and the fragmentation of quarks into a $\Lambda$. 
Starting from  the general definition of quark distributions  
and fragmentation functions, which explicitly 
incorporates  energy momentum conservation, we were able to show 
the following: 
\begin{itemize}
\item{} The hyperfine interaction 
responsible for the $\Delta$-$N$ and $\Sigma$-$\Lambda$ 
mass splitting leads to quark distributions  
and fragmentation functions which differ 
significantly from those based on 
$SU(3)$ and $SU(6)$ symmetries.  
\item{} The hyperfine interaction 
leads to two main qualitative predictions for Lambda quark
distributions and quark fragmentation functions.   
\subitem{} First, it  implies that the strange quark distribution 
in the $\Lambda$ 
and the strange quark fragmentation functions into a Lambda 
are much harder than the corresponding up and down 
quark distributions/fragmentation 
functions. 
\subitem{} Second, it predicts that the non-strange valence 
quarks of the $\Lambda$ are polarized and hence that non-strange 
quarks can fragment into polarized Lambda's.        
\item{} The relative magnitude of the 
non-strange quark polarization is substantial for large Bjorken-$x$ 
values, where both the polarized and the unpolarized quark 
distributions are governed by the scalar diquark component of the 
wave function. 
This large non-strange polarization will 
dominate any observable in which the strange component 
is suppressed. 
\item{} Our approach also gives  a natural explanation for 
the dependence of the maximum of the 
$\xi=\ln(1/z)$ spectrum on the type of particles produced  
in $e^+e^-$ annihilation.  
\end{itemize}

While all these associations follow quite naturally from the 
general definitions of the quark distributions 
and fragmentation functions and 
energy-momentum conservation, the magnitude of these effects 
has to be calculated in a model-dependent way. 
We calculated the quark distribution and fragmentation functions 
in the MIT bag model, using the Peierls-Yoccoz projection method 
to construct translationally invariant states. 
The calculated 
fragmentation functions give an overall good description of all 
measured observables and are in far better agreement with the data 
than flavor symmetric  models.    
We predict positive and significant $\Lambda$ polarization  
in semi-inclusive DIS experiments induced by charged leptons, 
while models based on SU(3) flavor symmetry 
predict zero or negative $\Lambda$ polarization.  

\acknowledgments

We would like to thank W. Melnitchouk, 
A.W. Schreiber  and K. Tsushima  
for helpful discussions. This work was partly supported
by the Australian Research Council.  One of the authors [JTL] 
was supported in part by National Science Foundation research
contract PHY-9722706.  One author [JTL] wishes to thank the
Special Research Centre for the Subatomic Structure of 
Matter for its hospitality during the time this work was 
carried out.

\begin{table} 
\caption{The probabilities for finding a quark polarized 
parallel ($\uparrow$) or anti-parallel ($\downarrow$) 
to the Lambda and accompanied by a scalar ($s$) or 
vector ($v$) diquark. } 
\label{tab:1} 
\begin{tabular}{|c|c|c|c|}
\rule[-0.4cm]{0cm}{1cm}
$q(qq)_v$ & $P[q(qq)_v]$ & 
$q(qq)_s$ & $P[q(qq)_s]$ 
\rule[-0.4cm]{0cm}{1cm}\\ 
\hline 
\rule[-0.4cm]{0cm}{1cm}
$u_v^\uparrow =d^\uparrow_v$ & $1/12$ & 
$u_s^\uparrow =d^\uparrow_s$ & $1/12$  
\rule[-0.4cm]{0cm}{1cm}\\ 
$u_v^\downarrow =d^\downarrow_v$ & $2/12$ & 
$u_s^\downarrow =d^\downarrow_s$ & $0$ 
\rule[-0.4cm]{0cm}{1cm}\\
$s_v^\uparrow$ & $0$ & 
$s_s^\uparrow$ & $4/12$ 
\rule[-0.4cm]{0cm}{1cm}\\
$s_v^\downarrow$ & $0$ & 
$s_s^\downarrow$ & $0$ 
\rule[-0.4cm]{0cm}{1cm}\\ 
\hline 
\rule[-0.4cm]{0cm}{1cm}
net & $1/2$ & net & $1/2$ \\ 
\end{tabular} 
\end{table}

\end{document}